\documentclass[aps,prb,twocolumn]{revtex4-1}
\usepackage{graphicx}
\usepackage{bm}
\usepackage{amsmath}

\newcommand{\g}{\ifmmode\text{g}\else{g}\fi}
\newcommand{\rrho}{ \mbox{\boldmath{$\rho$}}}

\begin{document}

\title{ Prediction of the spin triplet two-electron quantum dots in Si: towards controlled quantum simulations of magnetic systems}
	
\date{\today}

\author{D.~Miserev} 
\affiliation{Department of Physics, University of Basel, Klingelbergstrasse 82, CH-4056 Basel, Switzerland}

\author{O.~P.~Sushkov} 
\affiliation{School of Physics, University of New South Wales, Sydney, Australia}

\begin{abstract}
  Ground state of two-electron quantum dots in single-valley materials like GaAs is always a spin singlet regardless of what the potential and interactions are. 
  This statement cannot be generalized to the multi-valley   materials like $n$-doped Si. 
  Here we calculate numerically the spectrum of a two-electron Si quantum dot and show
  that the dot with the lateral size of several nm can have the spin triplet ground state which is impossible in the single-valley materials.
  Predicted singlet-triplet level crossing in two-electron Si quantum dots
  can potentially establish the platform for quantum simulation of magnetic many body systems based on quantum dots.
  We suggest several examples of such systems that open a way to controlled quantum simulations within the condensed matter setting.
\end{abstract} 
	
\maketitle

\section{Introduction}

Electrons in Si have the valley degree of freedom~\cite{Zwanenburg2013} which makes them qualitatively different from electrons in atoms or in one-valley materials. 
In this paper we concentrate on properties of two-electron bound states in Si.
In atomic physics the ground state of two bound electrons is always a spin singlet. This is a general property that is independent of potential well and interaction~\cite{ashcroft}. 
The proof of this statement is based on that the ground state wave function in one-valley materials must have no nodes.
This is no longer valid for multi-valley electrons such as electrons in Si, whose ground state wave functions can have arbitrary number of nodes. 
In this paper we predict the regime when the ground state of a tunable two-electron Si quantum dot is the spin triplet.

Tunable quantum dots are typically built in heterostructures where the potential along the $z$ axis is presented by the layer edges and is much stronger than the lateral $(x, y)$ potential that is controlled by the electrostatic gates.
Tunnelling between valleys lifts the valley degeneracy \cite{borselli,Goswami2007}.
Single-electron valley splitting $\omega_0$ is theoretically  well-understood tunneling problem~\cite{Saravia2011} that is very sensitive to the interface potential~\cite{Saraiva2009}.

The effect of Coulomb interaction in two electron Si quantum dots has been considered in previous works~\cite{Burmistrov,Jiang2013}.
In these works only the long-range part of the Coulomb interaction has been taken into account. 
The result of Ref.~\cite{Jiang2013} is extremely small while the estimate of Ref.~\cite{Burmistrov} is roughly consistent with our value of the long range contribution.
However, in both papers~\cite{Burmistrov,Jiang2013} the
on-site Hubbard repulsion is missing.
In the present work we calculate both the Hubbard and the long range contribution numerically. 
The Hubbard gives the leading contribution which results in a large value of the effective exchange.
Our result is important as the large value of the exchange makes a qualitative difference. 
In particular, we predict that in a quantum dot with the lateral size smaller than several nm the exchange interaction becomes larger than the single-electron valley splitting $\omega_0$
which experimental range is $\omega_0 \sim 0.1-1.5\,$meV, see Refs.~\cite{Ando1982,Goswami2007,Yang}.
This results in the spin-triplet ground state which is a qualitatively new phenomenon
compared to the conventional quantum dots~\cite{ashcroft}.
  
In this work we considered three different interface potentials.
The single-electron valley splitting $\omega_0$ is extremely sensitive to the interface potential in agreement with the previous research \cite{Saravia2011,Saraiva2009}.
At the same time, the Coulomb exchange matrix element is robust and insensitive to the interface potential and thus, to the quality of interface.

The singlet-triplet level crossing can be potentially driven by (i) the lateral confinement (size of the dot) or by (ii) the back gate. 
In the first case one varies the exchange Coulomb matrix element which is inversely proportional to the dot volume. 
In the second case one varies the single electron valley splitting $\omega_0$, see Ref.~\cite{Yang}.
Apart of the fundamental importance of the $S = 1$ ground state,
the tunable singlet-triplet level crossing can be used to create artificial multi-dot magnetic systems which exhibit
strongly correlated many body physics with externally driven quantum phase transitions. The spirit of this idea is similar to
quantum simulations of the Mott-Hubbard model in arrays of quantum
dots \cite{Singhal2011,Salfi2016,Hensgens2017}.
However, there are two important differences. 
(i) In contrast with the Mott-Hubbard case where there is no external handle to drive the quantum phase transition, here we have tunable singlet-triplet level crossing that allows to drive a quantum phase transition by electrostatic gates.
(ii) Absence of the charge dynamics in spin systems makes them almost insensitive to the Coulomb disorder~\cite{Tkachenko} which significantly reduces requirements to the quality of nanofabrication.

From the experimental point of view, the local spin measurements that are required for the artificial magnetic systems, can be already done with the help of spin-polarized scanning tunnelling microscopy \cite{Haze,Wiesendanger2018,Wiesendanger2019}.
As the potential artificial magnetic systems, we discuss Haldane spin 1 chain~\cite{Haldane}, the topological spin 1/2 edge states~\cite{Kennedy},
and the quantum phase transition from Haldane chain to the
``antiferromagnetic spin ladder''~\cite{Kotov}.
We also discuss $O(3)$ quantum criticality in square arrays~\cite{Shevchenko}
and underline quantum criticality in triangular arrays where
the nature of quantum phase transition
is not understood theoretically.

\section{Theoretical model}

We describe electron dispersion along the $z = [0 0 1]$ direction using one-dimensional (1D) tight binding model suggested
in Ref.~\cite{Boykin2004}.
In order to reproduce two degenerate minima of the Si dispersion along the $z$ direction, one has to account for the nearest and next-to-nearest neighbor hopping terms given by the matrix elements $v$ and $u$, respectively:
\begin{eqnarray}
H_z & = & \sum_{i_z, \sigma = \uparrow,\downarrow} \left\{v \, c^{\dag}_{i_z,\sigma} c_{i_z + 1,\sigma}
+ u \, c^{\dag}_{i_z,\sigma} c_{i_z + 2,\sigma} + h.c.\right.\nonumber\\
&+&
\left.\left(V(i_z) + \varepsilon_0 \right) c_{i_z,\sigma}^{\dag} c_{i_z,\sigma} \right\} .
\label{Hz}
\end{eqnarray}
Here $c_{i_z,\sigma}^{\dag}$ is the electron creation operator at the site $i_z$ with the spin projection $\sigma$, $V(i_z)$ is the heterostructure interface potential. 
The constant $\varepsilon_0$ is chosen to set the dispersion minimum at $\varepsilon = 0$.
Hopping parameters $v \approx 0.68 \,$eV, $u \approx 0.61 \,$eV fit the Si dispersion the best \cite{Boykin2004}.
The free electron spectrum given by the Hamiltonian (\ref{Hz}) is discussed in the Appendix.

In this work we consider the pancake geometry $d \ll D$ of a quantum dot, where $d$ is the heterostructure width and $D$ is the lateral size of the wave function that is defined by the electrostatic gates.
The quantum dot sizes $d$ and $D$ are defined through the inverse participation ratios:
\begin{eqnarray}
\frac{1}{D^2} = \int \phi^4(\rrho) \, d \rrho ,  \quad \frac{1}{d} = \int \Phi^4(z) \, dz .
\label{ipr}
\end{eqnarray}
Here $\Phi(z)$ is the wave function along the $z$ direction, $\phi(\rrho)$ is the lateral wave function, $\rrho = (x, y)$.
As the valley splitting $\omega_0$ is much smaller than the lateral level spacing, $\phi (\rrho)$ can be taken the same for the lowest valley split states.
The actual lateral wave function is not important and for numerical calculations we model it by the Gaussian.

Finally, we introduce the electron-electron interaction:
\begin{eqnarray}
H_C = U_H \sum_i c_{i\uparrow}^{\dag} c_{i\downarrow}^{\dag} c_{i\downarrow} c_{i\uparrow} 
+ \frac{1}{2} \sum_{\substack{i\ne j\\
    \alpha,\beta}}
V_{ij} c_{i\alpha}^{\dag} c_{j\beta}^{\dag} c_{j\beta} c_{i\alpha},
\label{CI}
 \end{eqnarray}
where $\alpha, \beta = \uparrow, \downarrow$ are spin indexes, $i = (i_x, i_y, i_z)$ enumerates the lattice sites, 
$U_H$ is the on-site Hubbard interaction and $V_{i j}$ is the long range Coulomb:
\begin{eqnarray}
V_{ij} = \frac{V_0}{\sqrt{(i_x - j_x)^2+ (i_y - j_y)^2 + (i_z - j_z)^2}}, \label{Vij}
\end{eqnarray}
where $V_0 = e^2 / \epsilon b$ is the electron-electron interaction at nearest sites, $e$ is the electron charge, $\epsilon$ is the dielectric constant, $b = 1.36\,$\AA \ is the inter-atomic scale. 
In this paper we use the Coulomb parameters calculated 
in Ref.~\cite{DFT} via DFT+U+V method: $U_H\approx 3.5 \,$eV, $V_0\approx 1.35 \,$eV.

We consider three different shapes of the interface potential $V(i_z)$: rectangular, parabolic and parabolic with $\delta$-doping. 
In the latter case we model the impurity-doped monolayer by a positively charged plane creating the potential $\delta V(i_z) = e \alpha  b  |i_z|$, where $\alpha$ is the electric field created by the charged plane. 
For further calculations we choose $\alpha = 15.4 \,$meV/nm.
All interface potentials are chosen such that the inverse participation ratio $d = 1 \,$nm in all cases, see Eq.~(\ref{ipr}).

The single-electron valley splitting $\omega_0$ is presented in Table~\ref{T1} and calculated as the difference between two lowest eigenstates of the Hamiltonian (\ref{Hz}).
As the smooth lateral potential defining the lateral size $D$ of the quantum dot does not affect the valley splitting $\omega_0$, $\omega_0$ is calculated in case of no lateral potential.
\begin{table}[h!]
  \caption{Energies of four lowest states of the Hamiltonian (\ref{Hz}) and the ground state
    splitting $\omega_0 = \varepsilon_2 - \varepsilon_1$ (given in meV) for different shapes of the interface potential. The size of the ground state in all cases is $d = 1 \,$nm.}
	\begin{tabular}{|l|l|l|l|l|l|}
		  \hline
	shape  & $\varepsilon_1$ & $\varepsilon_2$ & $\varepsilon_3$ & $\varepsilon_4$& $\omega_0$ \\
			\hline
	rectangular & 73.87 & 85.79 & 287.10 & 317.61& 11.93 \\ 
	parabolic  & 105.08 & 105.08 & 310.64 & 310.64 & $6\times 10^{-9}$\\
    $\delta$-doping &  110.22 & 110.52  &  321.18 &  321.21 & 0.30 \\
			\hline 
	\end{tabular}
    \label{T1}
\end{table}
The ground state splitting for the rectangular well is large,
$\omega_0 \sim 10 \,$meV, while for the parabolic well $\omega_0$ is practically zero. 
This illustrates the super-strong sensitivity of the
single-electron valley splitting $\omega_0$ to properties of the interface,
for sizable $\omega_0$ one needs a sharp interface~\cite{Saraiva2009,Boykin2004}. 
The moderate $\delta$-doping allows to obtain intermediate values of $\omega_0$ that are consistent with the experiments \cite{Ando1982,Goswami2007,Yang}. 
For example, the chosen value of $\alpha = 15.4 \,$meV/nm is sufficient to give $\omega_0 = 0.3 \,$meV.

\section{Low-energy spectrum of two-electron Si quantum dot}

\begin{figure}[h]
	\includegraphics[width = 0.5 \columnwidth]{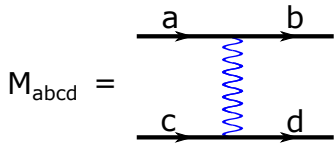}
	\caption{Matrix elements $M_{abcd}$, see Eq.~(\ref{M}), of the Coulomb interaction (\ref{CI}).
	}
	\label{diag}
\end{figure}

We start from calculating matrix elements of the Coulomb interaction $M_{abcd}$,
see Fig.~\ref{diag}:
\begin{eqnarray}
M_{abcd} = \sum\limits_{i, j} \Psi_b^*(i) \Psi_d^*(j) (U_H \delta_{ij} + V_{ij}) \Psi_c(j) \Psi_a(i) . 
\label{M}
\end{eqnarray}
Here $\Psi_a (i) = \phi(i_x, i_y) \Phi_a(i_z)$, $\phi$ is the 
Gaussian lateral function and $\Phi_a$ is the exact eigenfunction
of the Hamiltonian (\ref{Hz}).
Indexes $a, b, c, d \in \{1, 2\}$ label first two single electron states that are split by $\omega_0$, see Table~\ref{T1}.
Nonzero matrix elements for the quantum dot with $d = 1\,$nm and $D = 4 \,$nm are presented in Table~\ref{Tp}.
For these particular potentials
$M_{1112}=M_{1222}=0$ due to the symmetry $z \to -z$. 
We have also checked (not presented here) that
for the asymmetric potentials these matrix elements are still negligible.
\begin{table}[h!]
  \caption{Nonzero Coulomb matrix elements $M_{abcd}$ (meV) for 
    rectangular, parabolic and $\delta$-doped interface potentials. The dot lateral size is $D = 4\,$nm, the Si layer width $d = 1 \,$nm.
    We present the Hubbard, $U_H$, and the long-range Coulomb contributions, $V$, see Eq.~(\ref{CI}), as well as their sum.
    For direct matrix elements we also present their values with subtracted charging energy $U_C = M_{1122}$.
  }
\begin{tabular}{|l|l|l|l|l|}
\hline
   & $M_{1111}$ & $M_{2222}$ & $M_{1122}$ & $M_{1212}$ \\
   \hline
rectangular&&&&\\
$V$ & 115.92 &117.40  & 116.52 &0.26   \\ 
$U_H$ &0.52  & 0.56   &0.18    &0.18  \\ 
$V+U_H$ &116.44  &117.96  &116.70  &0.44   \\    
$V+U_H-U_C$ &-0.27  &1.26  &0  &   \\    
   \hline
parabolic&&&&\\
$V$ & 115.96 & 115.96 & 115.84 & 0.06 \\ 
$U_H$ & 0.54 & 0.54 & 0.17 & 0.17\\ 
$V+U_H$ & 116.50 & 116.50 & 116.01 & {\bf 0.23}\\ 
$V+U_H-U_C$ &{\bf 0.49}  &{\bf 0.49}  &{\bf 0}  &   \\    
\hline
$\delta$-doping&&&&\\
$V$ &116.25  &116.23  &116.12  &0.06     \\ 
$U_H$ &0.55  &0.55  &0.18  &0.18  \\ 
$V+U_H$ & 116.80  & 116.78  & 116.30  &{\bf 0.24}  \\ 
  $V+U_H-U_C$ &{\bf 0.50}  &{\bf 0.48}  &{\bf 0}  &   \\    
\hline 
		\end{tabular}
        \label{Tp}
\end{table}
Direct matrix elements $M_{1111}$, $M_{2222}$, and $ M_{1122}$
are large because they contain the charging energy
  $U_C\equiv M_{1122}\approx 116 \,$meV.
  The charging energy does not influence the level order,
  therefore in Table \ref{Tp} we also present values of the direct matrix
  elements with subtracted $U_C$.
As we see from Table~\ref{Tp}, the Coulomb matrix elements are
very weakly  sensitive to the potential shape unlike the
single-electron splitting $\omega_0$ which varies by several orders of magnitude, see Table~\ref{T1}.
Thus the Coulomb matrix elements are not sensitive to details of the interface.

\begin{figure}[h]
	\includegraphics[width = 0.9\columnwidth]{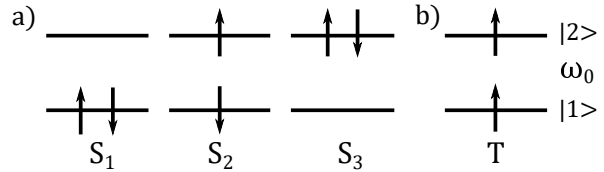}
	\caption{Lowest energy levels of a two-electron Si quantum dot: (a) three spin singlets, $S_1$, $S_2$, $S_3$; (b) the spin triplet $T$ (other two projections are not shown). Here $|1\rangle$, $|2 \rangle$ are two lowest one-particle states that are separated by the single-particle valley splitting $\omega_0$, see Table~\ref{T1}.
	}
	\label{levels}
\end{figure}
The low-energy spectrum of a two-electron Si quantum dot consists of three spin singlets, $S_1$, $S_2$, $S_3$ and one spin triplet, $T$, that are schematically shown in Fig.~\ref{levels}.
The singlet channel is described by the effective Hamiltonian
\begin{eqnarray}
H_S = \left(
\begin{array}{ccc}
M_{1111} - \omega_0 & 0 & M_{1212} \\
0 & M_{1122} + M_{1212} & 0\\
M_{1212} & 0 & M_{2222} + \omega_0 
\end{array}
\right) .
\label{hs1}
\end{eqnarray}
The triplet state energy reads:
\begin{eqnarray}
    \label{ht1}
    E_T = M_{1122} - M_{1212} .
\end{eqnarray}
At large single-electron valley splitting, $\omega_0 \gg 1 \,$meV,
the ground state is always $|S_1\rangle$.
This situation corresponds to the rectangular well, see Tables \ref{T1}, \ref{Tp}.
However, at $\omega_0 \lesssim 1$meV this is no longer true. 
After subtracting the charging energy $U_C$, $M_{1122} - U_C = 0$ and
$J \equiv M_{1111} - U_C \approx M_{2222}  - U_C \approx 2M_{1212} \approx 0.5 \,$meV,
see the bold font numbers in Table~\ref{Tp}.
Using this new notation $J$, we obtain the low-energy spectrum of a two-electron Si quantum dot:
\begin{eqnarray}
\label{el}
&&E_{\tilde{S}_1}=J-\sqrt{\omega_0^2+J^2/4} ,\nonumber\\
&&E_{S_2} = J/2 ,\nonumber\\
&&E_{\tilde{S}_3}=J+\sqrt{\omega_0^2+J^2/4} ,\nonumber\\
&&E_{T}=-J/2 .
\end{eqnarray}
Here $\tilde{S}_1$ and $\tilde{S}_3$ are linear combinations of singlets $S_1$ and $S_3$ diagonalizing the effective Hamiltonian (\ref{hs1}).
The spin triplet becomes the ground state when $E_T < E_{\tilde{S}_1}$ i.e. when $\omega_0 < \sqrt{2} J \approx 0.7 \,$meV.
Therefore, we get the triplet ground state for the $1 \times 4 \times 4 \,$nm quantum dot in case of the parabolic and $\delta$-doping interfaces.
This is very important result which is never possible in conventional two-electron quantum dots.

For the cases of parabolic and $\delta$-doping interface potentials where we predict the triplet ground state for the $1 \times 4 \times 4 \,$nm quantum dot, the exchange $J = 2 M_{1212} \approx 0.5 \,$meV is predominantly given by the Hubbard contribution, see the last row in Table \ref{Tp}. 
In the previous research \cite{Jiang2013,Burmistrov} only the long-range Coulomb has been taken into account which is not enough for any realistic setup.
Here we show explicitly that the triplet quantum dots are experimentally accessible due to the Hubbard contribution.

To measure the value of $J$ is not necessary to have small quantum dot $1 \times 4 \times 4 \,$nm. 
The Coulomb matrix element $J$ scales as $J \propto 1/(d D^2)$
~\cite{Burmistrov}. Our numerics confirms this scaling. Consider a larger dot
$1 \times 10 \times 10 \,$nm that is already technologically
available~\cite{Yang}. For such a dot the predicted exchange value is
$J\approx 0.08 \,$meV. Hence the ground state is the spin triplet if
$\omega_0 < \sqrt{2} J = 0.11 \,$meV.  However, even if $\omega_0$ is larger than 0.11 meV,
the value of $J$ can be measured as energy splitting between
$S_2$ and $T$ excited levels, see Eq.~(\ref{el}).

\section{Quantum simulators based on triplet quantum dots}
 
\begin{figure}[h]
	\includegraphics[width=0.35\textwidth]{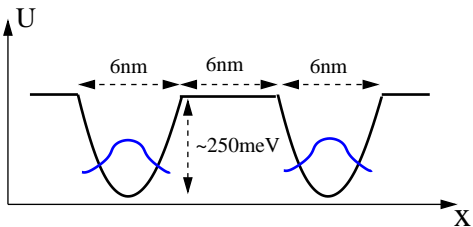}
	\caption{Schematic double dot setup.}
	\label{DW}
\end{figure}

Next, we propose to use Si quantum dots as building blocks for
simulation of quantum magnetic systems.
Quantum condensed matter simulators is new growing research direction \cite{Singhal2011,Salfi2016,Hensgens2017}. 
Due to the scaling $J \propto 1 / (d D^2)$ one can drive the singlet-triplet
level crossing within the dot by changing electrostatically the dot lateral size $D$. 
Alternatively, one can vary $\omega_0$ via the back gate voltage, see Ref.~\cite{Yang}.
Such electrostatic tunability allows to study driven quantum phase transitions,
a possibility that does not exist in previous quantum simulations proposals \cite{Singhal2011,Salfi2016,Hensgens2017}.
Consider a pair of $1 \times 4 \times 4 \,$nm quantum dots, see Fig.~\ref{DW}, the electron density is shown by blue lines. 
In absence of interaction the $1 \times 4 \times 4 \,$nm quantum dot 
requires the depth of the lateral potential $\sim 100 \,$meV. Taking into
account the charging energy $U_C = M_{1122} \approx 116 \,$meV, see Table~\ref{Tp}, the depth of quantum dot potential should be at least $200-250 \,$meV in order to accommodate two electrons. 

Assume that  the dot is in the spin triplet ground
state. Then there is an antiferromagnetic
interaction between spins ${\bf S}_{1,2}$ in different dots  due to the
Anderson superexchange \cite{anderson}:
\begin{eqnarray}
\label{j}
H_A = A \, {\bf S}_1 \cdot {\bf S}_2, \; A = \frac{4t^2}{U_C} \ ,
\end{eqnarray}
where $U_C\approx 116 \,$meV  is the charging energy, $t$ is the
tunnelling matrix element between the dots.
A simple estimate shows that the tunnelling matrix element is
$t \approx 5 \,$meV
when the distance between the dots $R = 10-15 \,$nm. In this situation
$A \sim 1 \,$meV. Value of $A$ is independently
tunable by the depth of the quantum well, see Fig.~\ref{DW}.
Variation of the depth from $250 \,$meV to $300 \,$meV is reducing $A$ by
an order of magnitude.

\begin{figure}[h]
\vspace{10pt}
	\includegraphics[width=0.4\textwidth]{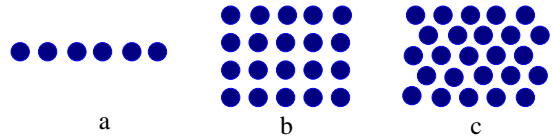}
	\caption{Linear (a), square (b), and triangular (c) lattices of
          quantum dots.}
	\label{ar}
      \end{figure}

Consider first the regime when the superexchange is smaller than the singlet-triplet splitting inside the dot, $A \ll E_T - E_{\tilde{S}_1} \sim 1 \,$meV.
Than we can disregard the dot spin singlet states and
build an array of antiferromagnetically interacting spins 1,
for example, a 1D array, Fig.~\ref{ar}(a).
This is already an interesting situation, the array represents  the Haldane spin chain~\cite{Haldane}, and hence, the topological spin $1/2$ edge states~\cite{Kennedy} can be observed experimentally.

The tunability makes the situation even more exciting.
Increasing $A$ or decreasing the in-dot singlet-triplet energy splitting,
one drives the multi-dot system to a quantum phase transition.
In the case of the Haldane chain this is the quantum phase transition to the
``antiferromagnetic spin ladder'' which allows to observe fractionalization
of spin excitations~\cite{Kotov}.

The 2D array of tunable dots
on square lattice, Fig.~\ref{ar}(b), manifests the $O(3)$ quantum critical physics, see e.g. Ref.~\cite{Shevchenko}.
This physics is well understood theoretically,
but experimentally it has been observed only in 3D
spin-dimerized compounds~\cite{Takatsu,Merchant}. The 2D $O(3)$
physics is different and it has never been observed experimentally.

The tunable 2D array on triangular lattice, Fig.~\ref{ar}(c), is especially
interesting. Deeply in the spin-triplet state
this is the spin 1 antiferromagnet on the triangular lattice. 
Driving the single dot singlet-triplet level crossing (changing the dot size)
would drive the multi-dot system to
the quantum disordered state. Such an experiment can shed the light on the
nature of this quantum phase transition which is not  understood theoretically.

In this paper we predict theoretically the singlet-triplet level crossing in two-electron Si quantum dots. This phenomenon is especially remarkable as such a level crossing is not possible in single-valley materials \cite{ashcroft}. The level crossing can be controlled by electrostatic gates. It opens opportunity for controlled quantum simulations of magnetic systems where single magnetic site is given by a triplet quantum dot. Here we suggest to simulate 1D spin systems (e.g. Haldane spin 1 chain hosting topological spin 1/2 edges states) and 2D arrays of quantum dots that are expected to have rich critical phenomena not yet observed experimentally.

\section{Acknowledgments}
We thank Dimitrie Culcer,
Susan Coppersmith, Andre Saraiva, Joe Salfi,
and Alexander Yaresko
for very important stimulating discussions.
D.M and O.P.S acknowledge support of the
Australian Research Council Centre of Excellence in Future
Low-Energy Electronics Technologies (project number
CE170100039) and funded by the Australian Government.
D.M acknowledges support of the Georg H. Endress foundation.

\appendix

\section{Folded and unfolded Brillouin zones}  

\begin{figure}[h]
	\includegraphics[width=0.49\textwidth]{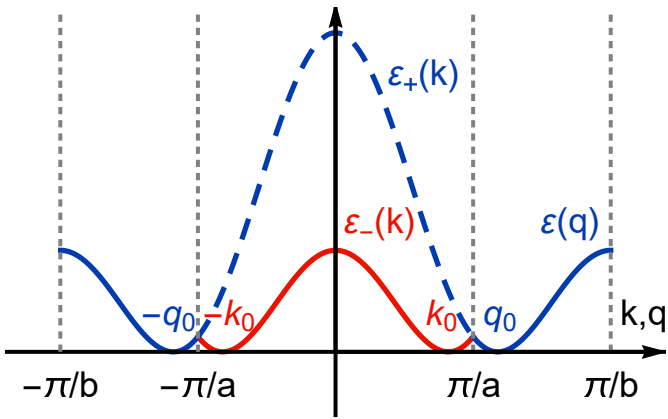}
	\caption{The ``unfolded'' and ``folded'' dispersions Eqs.~(\ref{unfold}), (\ref{disp}). Gray dotted lines show the ``unfolded'' $- \pi / b < q < \pi / b$ and ``folded'' $- \pi/ a < k < \pi / a$ Brillouin zones, $a = 2 b$. The branch $\varepsilon_+(k)$ coincides with the ``unfolded'' dispersion $\varepsilon (q)$ (indicated by the dashed blue line). The branch $\varepsilon_- (k)$ (the red line) is the mirror image of the rest of $\varepsilon (q)$ (the solid blue line). $\pm q_0$ ($\pm k_0$) are the dispersion minima corresponding to the ``unfolded'' (``folded'') description.
	}
	\label{dispF}
\end{figure}

Eigenstates of the Hamiltonian (\ref{Hz}) are plane waves:
\begin{equation}
|q \rangle = \frac{1}{\sqrt{2 N}} \sum \limits_{i_z} e^{i q b i_z} c_{i_z}^\dag |0 \rangle 
\end{equation}
with the spectrum 
\begin{eqnarray}
\varepsilon (q) = \varepsilon_0 + 2 v \cos(q b) + 2 u \cos(2 q b) , 
\label{unfold}
\end{eqnarray} 
where $-\pi / b < q < \pi / b$ is the first Brillouin zone (BZ), $b = 1.36 \,$\AA \ is the inter-atomic spacing, $2 N$ is the total number of Si atoms in the chain. Two minima of the dispersion correspond to $\pm q_0$, $q_0 = \arccos (- v / 4 u) /b \approx 0.59 \pi / b$. This is the ``unfolded'' description, see Fig.~\ref{dispF}.

\begin{figure}[h]
	\includegraphics[width=0.25\textwidth]{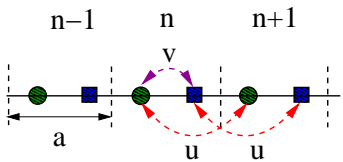}
	
	\caption{1D model for the two valley $z$-dispersion in Si
		interface~\cite{Boykin2004}. There are two Si atoms in the elementary cell. The lattice spacing is
		$a = 2 b = 2.7 \,$\AA, $v$ and $u$ are the nearest and next-to-nearest neighbor hoppings, respectively.}
	
	\label{lat}
\end{figure}

We could alternatively consider the Si chain with the elementary cell consisting of two atoms, such that new lattice spacing is $a = 2 b = 2.7 \,$\AA , see Fig.~\ref{lat}. 
The wave function that accounts for the unit cell structure is then the following:
\begin{eqnarray}
| k \rangle = \frac{1}{\sqrt{N}} \sum\limits_{i_z} e^{i k a i_z}\left(\alpha \xi^{\dag}_{i_z} + \beta \eta^{\dag}_{i_z} \right)| 0 \rangle ,
\label{wf}
\end{eqnarray}
where $\xi_{i_z}^{\dag} = c^\dag_{2 i_z}$ ($\eta_{i_z}^{\dag} = c^\dag_{2 i_z + 1}$) is the electron creation operator at the ``circle'' (``square'') cite of the lattice, see Fig.~\ref{lat}.  $N$ is the total number of unit cells in the lattice. 
The dispersion $\varepsilon_\pm(k)$ and the Bloch amplitudes $\mathcal{B}_\pm(k)$ are the following:
\begin{eqnarray}
&& \varepsilon_\pm (k) = \varepsilon_0 + 2 u \cos(ka) \pm  2 v \cos(ka/2) , \label{disp}\\
&& \mathcal{B}_\pm(k) ={\alpha \choose \beta} = \frac{1}{\sqrt{2}} {1 \choose \pm e^{i k a /2}} \ .
\label{bloch}
\end{eqnarray}
This is the ``folded'' description, see Fig.~\ref{dispF}. Here we use quasimomentum $k$ for the ``folded'' case,
$-\pi / a < k < \pi / a$ and 
$q$ for the ``unfolded'' case,  $-\pi / b < q < \pi / b$.

The lower branch of dispersion, see Fig.~\ref{dispF}, is given by $\varepsilon_-(k)$. 
Minima of $\varepsilon_-(k)$ are at $\pm k_0$, $k_0 = 2 \arccos (v / 4 u) /a \approx 0.82 \, \pi / a$. 
The corresponding wave function (\ref{wf}) can be written as
\begin{eqnarray}
\label{wfB}
\psi_- (z)  = e^{i k z} \mathcal{B}_- (k) \ ,
\end{eqnarray}
where $\mathcal{B}_- (k) $ is the Bloch amplitude, Eq.~(\ref{bloch}). The overlap of Bloch amplitudes corresponding to two minima of the dispersion is then the following:
\begin{eqnarray}
\label{ov} |\mathcal{B}_-^\dag (- k_0) \cdot \mathcal{B}_- (k_0) |= \cos \left(\frac{k_0 a}{2} \right) = \frac{v}{4 u} \approx 0.28 \ .
\end{eqnarray}

The overlap of Bloch functions corresponding to different valleys,
Eq.~(\ref{ov}), was largely underestimated in Ref.~\cite{Jiang2013}.
This is why their value of the long range exchange Coulomb matrix element is
by two orders of magnitude smaller than that of Ref.~\cite{Burmistrov}.
Both these works missed the Hubbard contribution to the exchange which is the dominant contribution to the effective exchange interaction.


\begin{thebibliography}{10}

\bibitem{Zwanenburg2013} F. A. Zwanenburg, A. S. Dzurak, A. Morello, M. Y. Simmons,
L. C. L. Hollenberg, G. Klimeck, S. Rogge, S. N. Coppersmith, and
M. A. Eriksson, Rev. Mod. Phys. {\bf 85}, 961 (2013).


\bibitem{ashcroft}  N. W. Ashcroft, and N. D. Mermin, \textit{Solid State Physics}, Saunders, New York, 1974.

\bibitem{borselli}  M. G. Borselli, R. S. Ross, A. A. Kiselev, E. T. Croke, K. S. Holabird, P. W. Deelman, L. D. Warren, I. Alvarado-Rodriguez, I. Milosavljevic, F. C. Ku, W. S. Wong, A. E. Schmitz, M. Sokolich, M. F. Gyure, and A. T. Hunter, Appl. Phys. Lett. \textbf{98}, 123118 (2011).

\bibitem{Goswami2007} S. Goswami, K. A. Slinker, M. Friesen, L. M. McGuire, J. L. Truitt, C. Tahan, L. J. Klein, J. O. Chu, P. M. Mooney, D. W. van der Weide, R. Joynt, S. N. Coppersmith, and M. A. Eriksson, Nat. Phys. {\bf 3}, 41 (2007).

\bibitem{Saravia2011} A. L. Saraiva, M. J. Calderon, R. B. Capaz, X. Hu, S. Das Sarma, and 
B. Koiller, Phys. Rev. B {\bf 84}, 155320 (2011).

\bibitem{Saraiva2009} A. L. Saraiva, M. J. Calderon, X. Hu, S. Das Sarma, and B. Koiller, Phys. Rev. B \textbf{80}, 081305(R) (2009).

\bibitem{Burmistrov} A. U. Sharafutdinov and I. S. Burmistrov, J. Phys.: Condens. Matter \textbf{24}, 155301 (2012).

\bibitem{Jiang2013} L. Jiang,  C. H. Yang,  Z. Pan,  A. Rossi,  A. S. Dzurak, and D. Culcer, Phys. Rev. B {\bf 88}, 085311 (2013).

\bibitem{com1} The exchange matrix element depends on the size of the dot. Of course, we compare dots of the same size.
  
\bibitem{Ando1982} T. Ando, A. B. Fowler, and F. Stern, Rev. Mod. Phys. {\bf 54}, 437 (1982).


\bibitem{Yang}  C. H. Yang, A. Rossi, R. Ruskov, N. S. Lai, F. A. Mohiyaddin, S. Lee, C. Tahan, G. Klimeck, A. Morello and A. S. Dzurak, Nat. Commun. \textbf{4}, 2069 (2013). 

\bibitem{Singhal2011} A. Singha, M. Gibertini, B. Karmakar, S. Yuan, M. Polini, G. Vignale, 
M. I. Katsnelson, A. Pinczuk, L. N. Pfeiffer, K. W. West, V. Pellegrini, Science {\bf 332}, 1176 (2011). 

\bibitem{Salfi2016}
J. Salfi, J. A. Mol, R. Rahman, G. Klimeck, M. Y. Simmons, L. C. L. Hollenberg, and S. Rogge,
Nat. Commun. {\bf 7}, 11342 (2016). 

\bibitem{Hensgens2017} T. Hensgens, T. Fujita, L. Janssen, X. Li, C. J. Van Diepen, C. Reichl, W. Wegscheider, S. Das Sarma, and  L. M. K. Vandersypen, Nature {\bf 548}, 70 (2017). 

\bibitem{Tkachenko} O. A. Tkachenko, V. A. Tkachenko, I. S. Terekhov, O. P. Sushkov, 2D Mater. {\bf 2}, 014010 (2015).



\bibitem{Haze} M. Haze, Hung-Hsiang Yang, K. Asakawa, N. Watanabe, R. Yamamoto,   Y. Yoshida, and   Y. Hasegawa, Rev. Sci. Instrum. \textbf{90}, 013704 (2019).

\bibitem{Wiesendanger2018} H. Kim, A. Palacio-Morales, T. Posske, L. Rózsa, K. Palotás, L. Szunyogh, M. Thorwart and R. Wiesendanger, Sci. Adv. \textbf{4},  eaar5251 (2018).

\bibitem{Wiesendanger2019} C. Friesen, H. Osterhage, J. Friedlein, A. Schlenhoff, R. Wiesendanger, S. Krause, Science \textbf{363}, 1065 (2019).

  
\bibitem{Haldane}
F.~D.~M.~Haldane, Phys. Lett. A {\bf 93}, 464 (1983); Phys. Rev. Lett. {\bf 50}, 1153 (1983).
\bibitem{Kennedy} T. Kennedy, J. Phys. Condens.: Matter  {\bf 2}, 5737 (1990).
\bibitem{Kotov} V. N. Kotov, O. P. Sushkov, R. Eder, 
Phys. Rev. B {\bf 59}, 6266 (1999).
\bibitem{Shevchenko} P. V. Shevchenko, A. W. Sandvik, O. P. Sushkov, Phys. Rev. B \textbf{61}, 3475 (2000).



\bibitem{Boykin2004}
  T. B. Boykin, G. Klimeck, M. A. Eriksson, M. Friesen, S. N. Coppersmith, P. von Allmen, F. Oyafuso, and S. Lee, Appl. Phys. Lett. {\bf 84}, 115 (2004).
 

\bibitem{DFT}V. L. Campo, Jr., and M. Cococcioni, J.  Phys. Condens. Matter {\bf 22}, 055602 (2010).


\bibitem{anderson} P. W. Anderson, Phys. Rev. \textbf{79}, 350 (1950).

\bibitem{Takatsu} K. Takatsu, W. Shiramura, and H. Tanaka, J. Phys. Soc. Jpn. \textbf{66}, 1611 (1997).
\bibitem{Merchant} P. Merchant, B. Normand, K. W. Krämer, M. Boehm, D. F. McMorrow, and Ch. R$\rm{ \ddot{u}}$egg, Nat. Phys. \textbf{10}, 373 (2014).




  
\end{thebibliography}
 \end{document}